 \documentclass[12pt,preprint]{aastex}
\slugcomment{Submitted 2003 May 22; Revised 2003 Aug 15}
\shorttitle{Oscillations in WZ Sge during Outburst}
\shortauthors{Welsh et al.}
\begin{document}

 \newcommand{\kms}{\mbox{$km \ s^{-1}$ }} 
 \newcommand{\Msun}{\mbox{M$_{\odot}$}}
 \newcommand{\Rsun}{\mbox{R$_{\odot}$ }}
 \newcommand{\Lsun}{\mbox{L$_{\odot}$ }}
 \newcommand{\ltsimeq}{\raisebox{-0.6ex}{$\,\stackrel
        {\raisebox{-.2ex}{$\textstyle <$}}{\sim}\,$}}
 \newcommand{\gtsimeq}{\raisebox{-0.6ex}{$\,\stackrel
        {\raisebox{-.2ex}{$\textstyle >$}}{\sim}\,$}}
%

\title{{\it Hubble Space Telescope} Observations of UV Oscillations in 
WZ~Sagittae During the Decline from Outburst\footnote{
Based on observations made with the NASA/ESA Hubble Space Telescope,
obtained at the Space Telescope Science Institute, which is operated by
the Association of Universities for Research in Astronomy, Inc., under
NASA contract NAS 5-26555.}
}

\author{William F. Welsh}
\affil{Department of Astronomy, San Diego State University,
San Diego, CA 92182-1221}
\email{wfw@sciences.sdsu.edu}

\author{Edward M. Sion}
\affil{Department of Astronomy and Astrophysics, Villanova 
University, Villanova, PA 19085}
\email{emsion@ast.villanova.edu}

\author{Patrick Godon\footnote{Visitor at the Space Telescope Science
Institute, 3700 San Martin Dr., Baltimore, MD 21218}}
\affil{Department of Astronomy and Astrophysics, Villanova University,
Villanova, PA 19085}
\email{pgodon@ast.vill.edu}

\author{Boris T. G\"ansicke}
\affil{Department of Physics and Astronomy, University of Southampton,
Southampton SO17 1BJ UK}
\email{btg@astro.soton.ac.uk}

\author{Christian Knigge}
\affil{Department of Physics and Astronomy, University of Southampton,
Southampton SO17 1BJ UK}
\email{knigge@astro.soton.ac.uk}

\author{Knox S. Long}
\affil{Space Telescope Science Institute, 3700 San Martin Drive,
Baltimore, MD 21218}
\email{long@stsci.edu}

\and

\author{Paula Szkody}
\affil{Department of Astronomy, University of Washington, Seattle, WA
98195}
\email{szkody@astro.washington.edu}


\begin{abstract}
We present a time series analysis of
{\it Hubble Space Telescope} observations of WZ~Sge 
obtained in 2001 September, October, November and December as WZ~Sge
declined from its 2001 July superoutburst. Previous analysis of these
data showed the temperature of the white dwarf decreased from
$\sim$29,000~K to $\sim$18,000~K.
In this study we binned the spectra over wavelength to yield ultraviolet
light curves at each epoch that were then analyzed for the presence of
the well--known 27.87~s and 28.96~s oscillations.  We detect the 29~s
periodicity at all four epochs, but the 28~s periodicity is absent. The
origin of these oscillations has been debated since their discovery in
the 1970s and competing hypotheses are based on either white dwarf
non--radial $g$--mode pulsations 
or magnetically--channelled accretion onto a rotating white dwarf.
By analogy with the ZZ Ceti stars, we argue that the 
non--radial $g$--mode pulsation model demands a strong dependence of
pulse period on
the white dwarf's temperature. However, these observations show the 29~s
oscillation is independent of the white dwarf's temperature. Thus we
reject the white dwarf non--radial $g$--mode pulsation hypothesis as the
sole origin of the oscillations. 
It remains unclear if magnetically--funnelled accretion onto a rapidly
rotating white dwarf (or belt on the white dwarf) is responsible for 
producing the oscillations. 
We also report the detection of a QPO with period $\sim$18~s in the
September light curve. The amplitudes of the 29~s oscillation and the QPO
vary erratically on short timescales and are not correlated with the mean
system brightness nor with each other.

\end{abstract}

\section{Introduction}
WZ~Sge is the prototype of a class of cataclysmic variables that
exhibit extreme characteristics: short orbital periods,
extremely large dwarf nova outbursts,
long outburst recurrence times, and low--mass companion stars.
WZ~Sge displays $\sim$7.5 mag outbursts separated by $\sim$2--3
decades and has a stellar mass ratio  of 13--25 : 1 with a
companion star that is less than 0.11 $\Msun$ and perhaps as low as 0.03
$\Msun$ (Steeghs et al.~2001, Ciardi et al.~1998).
These extreme characteristics make WZ~Sge an ideal proving
ground for accretion disk theory.

Among the many challenges WZ~Sge offers is the origin of the rapid 
27.87~s and 28.96~s oscillations. First seen in the optical by
Robinson, Nather \& Patterson (1978), these oscillations have been
detected in the UV (Welsh et al.~1997, Skidmore et al.~1999), infrared
(Skidmore et al.~2002) and in the X-ray (Patterson et al.~1998), though
weakly. The oscillations are complex, with large changes in amplitude,
phase jitter, and transient signals at nearby periods. Sometimes both
periodicities are present in the light curve.
Robinson et al.~(1978) proposed that the oscillations are due
to non--radial $g$--mode pulsations of the white dwarf, a hypothesis
supported by the simultaneous presence of the incommensurate 27.87~s and
28.96~s periodicities. Further support comes from analogy with GW~Lib, a
cataclysmic variable that almost certainly is a {\it bona fide} pulsating
white dwarf (van Zyl et al.~2000, Szkody et al.~2002).

An alternative hypothesis, proposed by Patterson (1980), attributed
the oscillations to a magnetized white dwarf channelling the accretion 
flow. The white dwarf's rotation provides the stable clock driving the
27.87~s fundamental periodicity. The other complex, transient
oscillations are due to reprocessing of radiation in the accretion disk
(see Patterson et al.~1998 for more details).
WZ~Sge would then be a member of the DQ~Herculis (or intermediate polar)
class of cataclysmic variables. The detection of a 27.87~s periodicity
in {\it ASCA} observations added much support to this magnetic accretor
model. However, despite the success of this ``oblique rotator'' model in
other cataclysmic variables, the presence of the simultaneous
incommensurate periodicities in WZ~Sge remains an obstacle for
this interpretation.

In their investigation of quasi--periodic oscillations in cataclysmic
variables, Warner \& Woudt (2002) developed the ``low inertia magnetic
accretor'' (LIMA) model and this yields a third possibility for the origin
of the oscillations. The model contains two elements:
(i) magnetically--controlled accretion onto a rapidly rotating belt on the
white dwarf and (ii) a prograde travelling wave at the inner edge of the
disk that produces a vertical thickening which acts as a site
for reprocessing and can also occult part of the disk and/or white
dwarf. In this scenario, the 27.87~s periodicity arises from magnetic
accretion onto the white dwarf,
but unlike the DQ~Her model, the belt can slip on the white dwarf's
surface and hence the oscillation can vary in phase, amplitude and even
period quite naturally. The 28.96~s signal comes from the reprocessing of
the 27.87~s signal by the travelling wave whose orbital period is 
$\sim$740~s. 
While the model can satisfactorally explain the periodicities, and perhaps
even the enigmatic $\sim$15~s quasi--periodic oscillation (QPO) 
discovered by Knigge et al.~(2002), no other dwarf nova shows persistent
QPOs in quiescence, requiring WZ~Sge to be unique in this regard.

Understanding the nature of the periodicities in WZ Sge is important not
only for its own sake, but because of the implications for 
accretion disk physics in general.
Due to its extreme outburst characteristics, WZ~Sge is
believed to have a low mass--transfer rate from its companion star
(e.g.~Osaki 1996). Yet this alone is insufficient to allow accretion disk
models to match the outburst size and timescale. One solution
requires that the accretion disk have a ``hole'' at its center (Lasota et
al.~1995).
This inner disk region may be evacuated by (i) a magnetic field, as in the
DQ~Her stars where the white dwarf's magnetic field
truncates the inner disk (Warner et al.~1996; Lasota et al.~1999);
or (ii) evaporation via a  ``siphon'' into a hot corona 
(Meyer \& Meyer--Hofmeister 1994). 
This coronal siphon mechanism has properties related to ADAF mechanisms 
in black-hole accreting systems (e.g.~Mineshige et al.~1998,
Meyer--Hofmeister \& Meyer~2001). 
Another solution requires the quiescent viscosity in WZ~Sge's
accretion disk to be very low: $\alpha_{cold} \ltsimeq 0.001$, compared
to a more typical value of $\alpha_{cold} \sim0.03$ in ``normal'' accretion
disks (e.g.~Smak 1993; Osaki 1996; Meyer--Hofmeister, Meyer \& Liu 1998). 
If WZ~Sge can be shown to be a DQ~Her--like system, then the 
``universality'' of the viscosity parameter is preserved and no seemingly
{\it ad hoc} low value for $\alpha_{cold}$ is necessary. On the other
hand, if WZ~Sge is a white dwarf pulsator, then either the coronal siphon
must be operating and we have a white dwarf analog of a black hole
accretion disk, or an anomalously low viscosity is required. Either way,
understanding the nature of the oscillations in WZ~Sge can have an
important impact on our understanding of accretion disks.

In 2001 July WZ~Sge went into outburst, 23 years after its previous
outburst (see Patterson et al.~2002 for extensive coverage of the
2001 outburst). Sion et al.~(2003) obtained {\it Hubble Space Telescope
(HST)} observations of WZ~Sge on the decline from outburst and in this
paper we present a time series analysis of those same data. 
These observations provide compelling evidence against a simple white
dwarf non--radial $g$--mode pulsation interpretation of the oscillations.

\section{The HST Observations} 
As part of the Director's Discretionary time, WZ Sge was observed with
the {\it HST} at four epochs:
2001 September 11, October 10, November 10, and December 11. 
Each observation lasted approximately 39 min, just under half
of the 81.6 min binary orbital period.
The UV light was probably dominated by emission from the heated white
dwarf from October to December; in September the disk contribution 
may have been substantial and spectral decomposition into disk and white
dwarf emission is much less certain (Sion et al.~2003).
See figure 1 in Long et al. (2003) for the outburst light curve and
the placement of the September, October and November HST observations.

The {\it HST} {{\it STIS FUV MAMA}} detector was used in {\it{TIME-TAG}}
mode with a wavelength coverage of 1140 -- 1735 \AA\ (see Sion et
al.~2003 for additional information regarding the observations).
Data reduction was carried out with the standard {\mbox{STSDAS}} pipeline
software. The data were cast into 2~s or 5~s time bins then binned 
over wavelength into a single UV light curve at each epoch.

The September observation took place just prior to the minimum before
the final re--brightening (or ``echo outburst'') of WZ Sge's slow and
erratic decline.
The light curve shows large amplitude rapid flickering; a rms
variation of 10.5\% of the mean flux (21.8~mJy) was present. 
In contrast, the October, November and December light curves show much
less flickering, with variations only slightly above the noise level. 
Figure~1 shows the four light curves, with respect to the start time of
each observation. The mean UV flux dropped by a factor of 2.4 between the
September and December observations.

\section{Power Density Spectra}
Figure 2 shows the power spectrum at each epoch, after prewhitening the
light curves with a sliding boxcar average of width $\sim$146~s. 
Since the duration of the light curves were short and therefore the
frequency resolution poor ($4.292 \times 10^{-4}$ Hz, or formally
equivalent to a period resolution of only 0.36~s at 28.96~s),
the power spectra were oversampled by a factor of 2, 
so only every other bin is independent. The power density 
is normalized such that a sinusoid of amplitude $A$ gives a power 
of $\frac{1}{2} A^{2}$. In all four cases, a spike consistent with the
previously known 28.96~s periodicity is present. Henceforth we shall
refer to this feature as the P29 signal. 
The period and amplitude of the P29 signal were determined via 
least--squares sine fitting.
Table 1 lists the observed characteristics of the P29 signal.

The luminosity of the P29 signal dropped by a factor $\sim$4.5 between
September and October, but thereafter remained fairly constant. In flux
units,  the amplitude changed from $\sim$0.42~mJy in September to
$\sim$0.09~mJy in October--December, while the fractional amplitude
dropped from $\sim$1.9\% to $\sim$0.9\%. 

The P29 luminosity experienced a larger decrease than the mean UV
luminosity, suggesting that P29 responds more quickly to changes in the
global mass accretion rate than the mean system. One possible explanation
is that the long cooling time of the white dwarf keeps the mean
system luminosity  well above its quiescent value; indeed in December 
WZ~Sge was still much brighter in the UV than when in quiescence.

The 27.87~s periodicity was not detected, nor was the $\sim$15~s
quasi--periodicity seen by Knigge et al.~(2002) in {\it HST} UV
observations earlier in the outburst. However, a broad feature near 18~s
is present in the September observations which we identify as a
quasi--periodic oscillation (QPO).

In Fig.~3 we show the September observations in more detail. The power
spectrum was recomputed using only a linear detrending, thus the rapid
rise in power at low-frequencies due to flickering remains present. 
Fig.~3 also shows the mean light curve and the amplitude (from the square
root of the power) of the P29 and QPO periodicities. The P29 and QPO
amplitudes were computed from short segments of the light curve, each
118~s in duration. Fifty such segments were used; they are not
independent, having 61\% overlap between them. The frequency
resolution of such short light curves is very poor 
so the power spectrum was oversampled by a factor of 4.
After removing the white noise background, the power was summed
over the intervals 27.8--33.7~s for the P29 signal and 16.9--19.7~s for
the QPO signal. 
The flickering red noise background remains present in these bandpasses,
so these amplitudes are still somewhat biased high;
the uncertainties in the amplitudes are roughly half the values of
amplitudes themselves.
The most noticeable features of the amplitudes are their rapid variations
and lack of correlation with the mean light curve or each other. No
periodicity in P29 or QPO amplitude is apparent. It is
interesting to note that despite the P29 signal being present at 4 epochs
spanning 3 months, the signal can vary from being quite strong to
undetectable in a span of less than a few hundred seconds. Long--term
stability is present while short--term stability is not. Although
hard to quantify, the amplitude of the QPO is roughly 0.31~mJy or 
1.4\% of the mean system brightness. If only a single sinusoid is
fit to the broad QPO signal, an amplitude of 0.27~mJy (1.3\%) 
at a period of 17.95~s is measured. 
The relationship, if any, between this weak QPO and the much stronger
$\sim$15~s and $\sim$6.5~s QPOs seen 20 days earlier in the outburst
(Knigge at al.~2002) is unclear.

\section{Discussion}
\subsection{The White Dwarf on the Decline from Outburst}
Based on the same {\it HST} UV spectra as used in this work, 
Sion et al.~(2003) measured the white dwarf's temperature decline from
29,000~K in 2001 September to 18,000~K in December.
Although some of the lines used in the analysis seem unusually narrow,
the interpretation of a cooling white dwarf is fairly secure based 
on the success of the models in matching the observations, in particular
the continuum and broad Ly $\alpha$ absorption.
Further evidence of an elevated white dwarf temperature comes from
the lack of strong quasi--molecular H$_{2}$ or C~I absorption lines that
are prominent in quiescence.
Nevertheless, the interpretation that the white dwarf completely
dominates the UV spectrum at these epochs is not fully established. 
If we omit the September observations on the presumption
that the accretion disk contributes significantly to this UV spectrum
there still remains a 5000 K drop in white dwarf temperature between
October and December.
The uncertainties in these temperature estimates are roughly 10\%, 
though the errors are largely systematic rather than statistical.

A similar change in white dwarf temperature was deduced from analysis of
{\it Far Ultraviolet Spectroscopic Explorer (FUSE)} observation by Long
et al.~(2003), with temperatures dropping from $\sim$25,000~K to
$\sim$22,600~K between September 29 and November 8.  Part of the
difference in estimated white dwarf temperature between the
non--simultaneous {\it HST} and {\it FUSE} spectra comes from the
uncertainty in the contribution of the disk to the total spectrum.
Nevertheless the trend --- a substantial cooling of the white
dwarf --- seems secure. 
Post--outburst cooling of the white dwarf has been well-documented in 
other dwarf novae, e.g., in VW~Hyi (G\"ansicke \& Beuermann 1996, Sion et
al.~1995), U~Gem (Sion et al.~1998), and AL~Com (Szkody et al.~1998).

Because of its critical importance in the interpretation to come, we
repeat the statement in Godon et al.~(2003) regarding the temperature of
the white dwarf in WZ~Sge. The temperature increase of the white dwarf,
due to the enhanced accretion during outburst, is not confined to a thin
outermost layer. Such a layer would very quickly radiate away heat gained
during the outburst and therefore would remain at an elevated temperature
only while substantial accretion is present. However, compressional
heating takes place deeper in the star (though still in the outer
layers) and takes months to cool, as is the case in WZ~Sge
(Sparks et al.~1993). 
So while irradiation from the boundary layer and inner disk heats the
surface of the white dwarf, compression gets under the white dwarf's
skin (Sion~1995, 1999).
Thus the dwarf nova outburst accretion
episode, which dumps on the order of $\sim 10^{-9} \ \Msun$ of
matter onto the white dwarf (Godon et al.~2003), produces a deep and
significant heating of the white dwarf's envelope.


\subsection{The White Dwarf Non--radial $g$--mode Pulsation Hypothesis}
Before we present the primary conclusion of our investigation, we
briefly summarize the plausibility of the pulsation hypothesis.

The original, and in some ways still best, argument in favor of the
white dwarf pulsation hypothesis is the observed simultaneous presence of
the incommensurate P28 and P29 periodicities (Robinson et al.~1978).
Incommensurate periodicities are common for the ZZ Ceti class of stars
(isolated non--radial $g$--mode pulsating white dwarfs; see Kepler \&
Bradley 1995 for a good review of white dwarfs and pulsation).
Also, the
white dwarf in GW~Lib is known to exhibit ZZ~Ceti--type pulsations (van
Zyl et al.~2000; Szkody et al.~2002). Additional support, albeit weak,
comes from the UV spectrum of the oscillations that are consistent with
a white dwarf photospheric origin (Skidmore et al.~1999, Welsh et
al.~1997).

Yet the ZZ Ceti stars (and GW~Lib) have much longer periods (hundreds
of seconds) than WZ Sge. Furthermore, the quiescent temperature of the
white dwarf in WZ~Sge, $\sim$14,600~K (Cheng et al.~1997, Godon et
al.~2003), exceeds the nominal upper temperature limit 
(the ``blue edge'') of $\sim$13,500~K of the instability strip for ZZ
Ceti stars. 
However, more massive white dwarfs have a hotter ``blue edge'' than the
canonical 0.6 $\Msun$ white dwarf (Bradley \& Winget 1994; Bergeron
et al.~1995; Giovannini et al.~1998) and recent studies do in fact
suggest WZ~Sge has a massive white dwarf, exceeding $\sim$0.8 $\Msun$ and
perhaps even as high as 1.2 $\Msun$ (Steeghs et al.~2001;  Skidmore et
al.~2000; Knigge et al.~2002; Long et al.~2003; Godon et al.~2003).
By linearly extrapolating the blue edge temperature versus mass presented
in Giovanninni et al.~1998, a $g$--mode pulsating white dwarf with a
temperature of
14,600~K would require a white dwarf mass $\gtsimeq1.2 \Msun$, barely
within acceptable mass estimates. However, the extrapolation to
values above $1 \Msun$ is highly uncertain so this crude mass limit should
be taken as nothing more than a statement that $g$--mode pulsations
cannot be ruled
out by quiescent white dwarf temperature arguments. Furthermore, as was 
suggested by Szkody et al.~(2002) for GW~Lib, it could be that the 
temperature determined for WZ~Sge is a global average, and a cooler 
region could be responsible for the pulsations --- for GW~Lib, a 
two--temperature white dwarf model fits substantially better then a one 
temperature model.
Note that the thickness of the surface convective zone responsible for
the pulsations in WZ~Sge should be very thin compared to other ZZ~Ceti
stars since the timescale of the periodicity depends on the depth of the
base of the convective zone (e.g.~see Bergeron et al.~1995) --- fast
periodicities require a shallow convective zone.

Skidmore et al.~(1999) extrapolated the observed temperature--period
relation of ZZ~Ceti stars using the data in Clemens (1993) and found the
extrapolation falls tantalizingly close to the quiescent temperature and
period of WZ~Sge. This bolstered the viability of the white dwarf
pulsation hypothesis, and mitigated the objection that the 28~s and 29~s
periods are too short to be related to the ZZ Ceti phenomenon. 
We re-examined this extrapolation of figure 2 of Clemens (1993) and while
the {\em observed} temperature--period relation for ZZ Ceti stars does
project close to WZ~Sge, the {\em theoretical} relation makes a dramatic
drop to very fast periodicities at temperatures just higher than the
observed ZZ~Ceti star temperatures. From figure 3 of Clemens (1993), for a
temperature of 14,600~K the thermal timescale at the bottom of the
ionization zone where the pulsations are generated (which should be
approximately equal to the observed periodicity) is on the order of
only 0.01~s. There are large uncertainties in both the theory and
observed temperatures, so this 0.01~s timescale should not be
overinterpreted, but the conclusion remains thus: extrapolation of the
observed temperature--period relation of ZZ~Ceti stars to the WZ~Sge
regime cannot be trusted.
However, we can be confident that if WZ~Sge is a
non--radial $g$--mode pulsator, it should follow, at least crudely, the
{\em trend} of the temperature--pulse period relation for isolated white
dwarf pulsators:  hotter surface temperatures stars have shorter periods
and smaller pulsation amplitudes (Clemens 1993). 

Given the significant and deep heating of the white dwarf's envelope
discussed in \S4.1, it would be astounding if the periods of any
non-radial $g$--mode pulsations in WZ~Sge did not change substantially as
the white dwarf's temperature changed. Yet in this work we
have shown the continued presence of the 29~s periodicity while the white
dwarf cools by over 10,000 K. During the outburst decline,
WZ~Sge no longer lies near the extrapolated
empirical ZZ~Ceti period--temperature relation and, more importantly, 
{\em the observed period of the P29 oscillation is independent of the
white dwarf's temperature.}  These observations therefore rule out
non--radial $g$--mode pulsations of the white dwarf as the {\em sole}
origin of both the 28~s and 29~s pulsations.


\subsection{Implications for Magnetic Accretor Hypotheses}
Having claimed the observations of the P29 oscillation herald the end
of the simple non--radial $g$--mode pulsation hypothesis, we are now
left with a challenge: Can the observations be interpreted within the
framework of the magnetically--channelled accretion flow models?
 
Patterson (1980) proposed the magnetic oblique--rotator model
for WZ~Sge (i.e.~the DQ~Her--type scenario) based on the stability of 
the 27.87~s oscillation and the ability to match some of the complex 
sideband structure in the power spectrum (the ``satellite periods'') 
with simple integer multiples of the orbital frequency. In this model 
the rotation of the white dwarf provides a stable clock that produces 
the primary 27.87~s oscillation while the accretion disk provides
(transient) structures that reprocess the primary signal. 
Cheng et al.~(1997) measured a rapidly rotating white dwarf
($V_{rot} \sin i = 1200^{+300}_{-400}$ km s$^{-1}$ in quiescence)
supporting the magnetic rotator model, though this rotational velocity
has not been confirmed with post--outburst observations. 
As Patterson et al.~1998 showed, this fast rotation requires a high--mass
(0.8--1.1$\Msun$) white dwarf  if the period of rotation of the white
dwarf is 28~s. Finally, the presence of a periodicity at 27.86~s in 
{\it ASCA} 2--6~keV light curves, along with a weaker signal in the
0.4--2~ keV band (one half cycle out of phase), lends considerable
support for the DQ~Her interpretation (Patterson et al.~1998). 
Lasota et al.~(1999) expand upon the DQ~Her model and suggest that
the rapidly spinning white dwarf actually ejects most of the matter
transferred from the secondary star via a magnetic propeller. The
P29 signal is caused by reprocessing of the P28 signal by blobs at the
{\em outer} edge of the disk. A prediction of the model is that the white
dwarf cannot be massive ($\gtsimeq 1$ \Msun), potentially a serious
problem if recent estimates suggesting a massive white dwarf are correct.

The LIMA model of Warner \& Woudt (2002) is similar in spirit to the
DQ~Her model of Patterson (1980), 
but has added flexibility: the periodicities are not tied to the white
dwarf's rotation, but instead to a thin belt on the white dwarf. 
Since the belt contains much less inertia than the white dwarf,
significant short timescale changes in rotation rate are possible.
In this model the belt itself amplifies the white dwarf's
magnetic field, allowing a white dwarf with very low magnetic field
strength to channel the accretion flow. One particularly attractive
feature of the model is that the vertical thickening at the inner disk
can also explain the ``dips'' seen in WZ~Sge's light curve. 
While the model seems very promising for the dwarf nova oscillations
(DNOs) and QPOs seen in other cataclysmic variables, WZ~Sge presents
more of a challenge. For all other systems the DNOs are present only in
outburst,\footnote{at least in the optical; Warner \& Woudt (2002)
point out an X-ray QPO seen during quiescence in HT Cas.}
while in WZ~Sge the DNO signal, P28, is often present in quiescence and
seems to have disappeared during the outburst. 


\subsubsection{The 28s Periodicity}
The P28 signal is not present in outburst (e.g. Patterson et al.~1981;
Knigge et al.~2002; this work), and remained hidden for nearly 18 years
after the 1978 outburst.  In the standard DQ~Her scenario, the lack of
the P28 oscillation during outburst can be explained by the greatly
enhanced mass accretion rate (see Patterson et al.~1998): the ram
pressure of the accreting gas overwhelms the magnetic pressure and
effectively crushes the magnetosphere back onto the surface of the 
white dwarf and accretion takes place along the equator, not at
the magnetic poles. The magnetic field is too weak to channel the
accretion flow during outburst, hence the photometric modulations vanish.

Even when deep in quiescence, observations do not always show the P28
signal; this is a puzzle. When present, the P28 signal maintains
long--timescale clock--like stability, though short--term phase and
amplitude changes are frequent. Inhomogeneities in the disk and or
accretion flow must be such that the P28 signal is either not produced,
hidden, or canceled for extended durations of time. 
Patterson et al.~(1998) claim that sometimes when the P28 signal 
was absent other nearby periodicities that could be identified as 
sidebands of the P28 signal were observed, suggesting the
P28 signal was present, just not detected. 
While the prolonged absence of the P28 signal after the 1978 outburst is
not quite fully understood, a plausible reason exists, and the stability,
sidelobes, and X--ray modulation can be interpreted, at least
consistently, within the context of the DQ~Her model.


\subsubsection{The 29s Periodicity}
As seen here and many times previously, the P29 signal can exhibit rapid
changes in amplitude and phase on timescales as short as minutes
(e.g.~Skidmore et al.~1997), yet the mean period is constant and on
occasion the signal remains quite stable --- perhaps even retaining phase
coherence for up to a week (Patterson et al.~1998). 
The P29 signal exists throughout most of the dwarf nova eruption cycle: 
before, during and after the superoutburst --- though perhaps not close
to the peak of outburst (Knigge et al.~2002).  The P29 signal does not
correlate with the instantaneous luminosity of the system (and hence
the inferred instantaneous $\dot{M}$ --- see Fig.~3), though its
amplitude is somewhat tied with the average brightness of the system on
the decay from outburst. As we have shown in this work, P29 is de-coupled
from the white dwarf temperature. 

As Patterson (1998) stated, the P29 signal does not naturally appear as
any known simple ratio or sum of periodicities in the system (orbital or
spin). In the DQ~Her model, the P29 signal is created by the
reprocessing of the P28 signal by some structure in the disk: P29
is the ``beat'' between P28 and a blob orbiting in the disk with a period
of $\sim$740~s, most likely at the inner edge of the disk. 
Similarly, in the LIMA model of Warner \& Woudt (2002), the P28 signal
corresponds to a DNO while the P29 signal corresponds to a QPO, created
by the reprocessing of the DNO signal. While a periodicity of 741~s has
been detected in WZ~Sge's light curve (Warner \& Woudt 2002), it
is very rare, and the DNO driving signal (i.e.~P28) was not present
at that time.

In the DQ~Her model, the P28 period remains constant during outburst
since it is maintained by the rotation of the entire white dwarf. 
In contrast, on the rise to outburst the period of the P29 signal 
should
increase 
as the radius of the reprocessing inner disk shrinks,
and the signal disappear entirely when the disk reaches the white dwarf.
As the disk hole (i.e.~magnetosphere) re--opens during the outburst
decline, the P29 oscillation should re--appear and decrease back to its
quiescent period. 

In the LIMA model, the DNO period should decrease as more and more 
angular momentum is accreted into the belt during the rise to outburst. 
The P29 (QPO) signal would also experience a period change,
for two reasons: it is being driven by the faster DNO and the inner
radius of the disk where reprocessing is occurring is shrinking. On the
decline from outburst, the periods should increase. A strong
period--luminosity relationship through the outburst should exist,
as is the case for DNOs of other systems (see Woudt \& Warner 2002 and
references therein).

In both magnetic accretor scenarios, the P29 signal should vary during
the outburst. Yet as we have shown here, this is not the case in WZ~Sge:
the P29 signal remains constant in period on the decline from outburst,
and with the same period it has in quiescence.
In fact, just the presence of the P29 signal in the absence of the
P28 signal during the outburst decay presents a significant problem for
the magnetic accretor models. In the DQ~Her model, if the magnetosphere
is crushed onto the white dwarf thereby quenching the P28 signal, the
inner edge of the disk is now coincident with the white dwarf's
surface. So not only is there no inner disk region to reprocess any
modulation, there is no modulation to reprocess. And yet the P29 signal
is quite strong, stronger in fact than in quiescence, at least in the
2001 September observations. As in the DQ~Her model, the LIMA model
requires the presence of the DNO at 28~s to drive the QPO at 29~s.
The complete absence of the P28 signal in the observations presented here 
is vexing. We clearly do not understand the mechanism producing the 
P29 signal.

The continued presence of the P29 signal during the outburst decay, 
as the disk reforms and the mass accretion rate and white dwarf
temperature dramatically change, suggests that a clock of exceptional
(long--term) stability is involved. The most immediate candidate is the
rotation of the white dwarf, but how this could work is not at all
apparent. We conclude that while the rotating magnetic accretor models
remain viable, they are sorely incomplete.

\section{Summary}

We present ultraviolet light curves of the dwarf nova WZ~Sge 
as it declined from its 2001 July superoutburst. The observations were
made with the {\it HST} STIS and were obtained on 2001 September,
October, November and December. The well--known 28.96~s periodicity was
present at all four epochs, but the 27.87~s periodicity was absent.
We also note the appearance of a QPO with period $\sim$18~s in
the September light curve.

Between 2001 September and December the temperature of the white dwarf
dropped from $\sim$30,000~K to $\sim$18,000~K (Sion et al.~2003).
We argue that by analogy with the ZZ~Ceti stars, the white dwarf
non--radial $g$--mode pulsation model for the origin of the 27.87~s and
28.96~s periodicities
(Robinson et al.~1976) demands a large change in the pulsation period as
the white dwarf's temperature changes. However, these observations do not
support this prediction: the 29~s oscillation period is observed to be
independent of the white dwarf's temperature. 
We consider this the {\it coup de grace} of the non--radial
$g$--mode pulsation hypothesis as the {\em sole} origin of both the 28~s
and 29~s pulsations.

In light of these observations, the alternative hypotheses for the 
origin of the periodicities, the venerable DQ~Her magnetic rotator 
model (Patterson 1978) and the LIMA model of Warner \& Woudt (2002),
do not fare particularly well either. In these models the 28~s
periodicity is related to rotation of either the entire white
dwarf or an accretion belt on the white dwarf. The 28~s modulation
is ultimately responsible for driving all the other periodicities
including the 29~s modulation. The presence of a relatively strong 
29~s periodicity in the total absence of the 28~s periodicity is a
problem for the models. 
While the models remain viable, they must be considered incomplete.
Finally, we remark that while non--radial $g$--mode pulsations cannot be
responsible for the oscillations in WZ~Sge, this does not necessarily
imply that other modes of white dwarf pulsation are ruled out. In
particular, the periods of non--radial $r$--modes (Papaloizou \& Pringle
1978) can be near those of WZ Sge and are not very temperature sensitive
so they may play a role in WZ~Sge's puzzling behavior.

\acknowledgments
This work was supported by NASA through grants GO-09304 and GO-09459
from the Space Telescope Science Institute, which is operated by the
Association of Universities for Research in Astronomy, Inc., under NASA
Contract NAS 5-26555. Support was also provided in part by NSF grant
99-01195, NASA ADP grant NAG5-8388 (EMS), and from NASA through STScI
grant HST-GO-08156 (WFW). BTG acknowledges support from a PPARC Advanced
Fellowship.


\clearpage 
\begin{figure}
\figurenum{1}
\epsscale{0.8}
\plotone{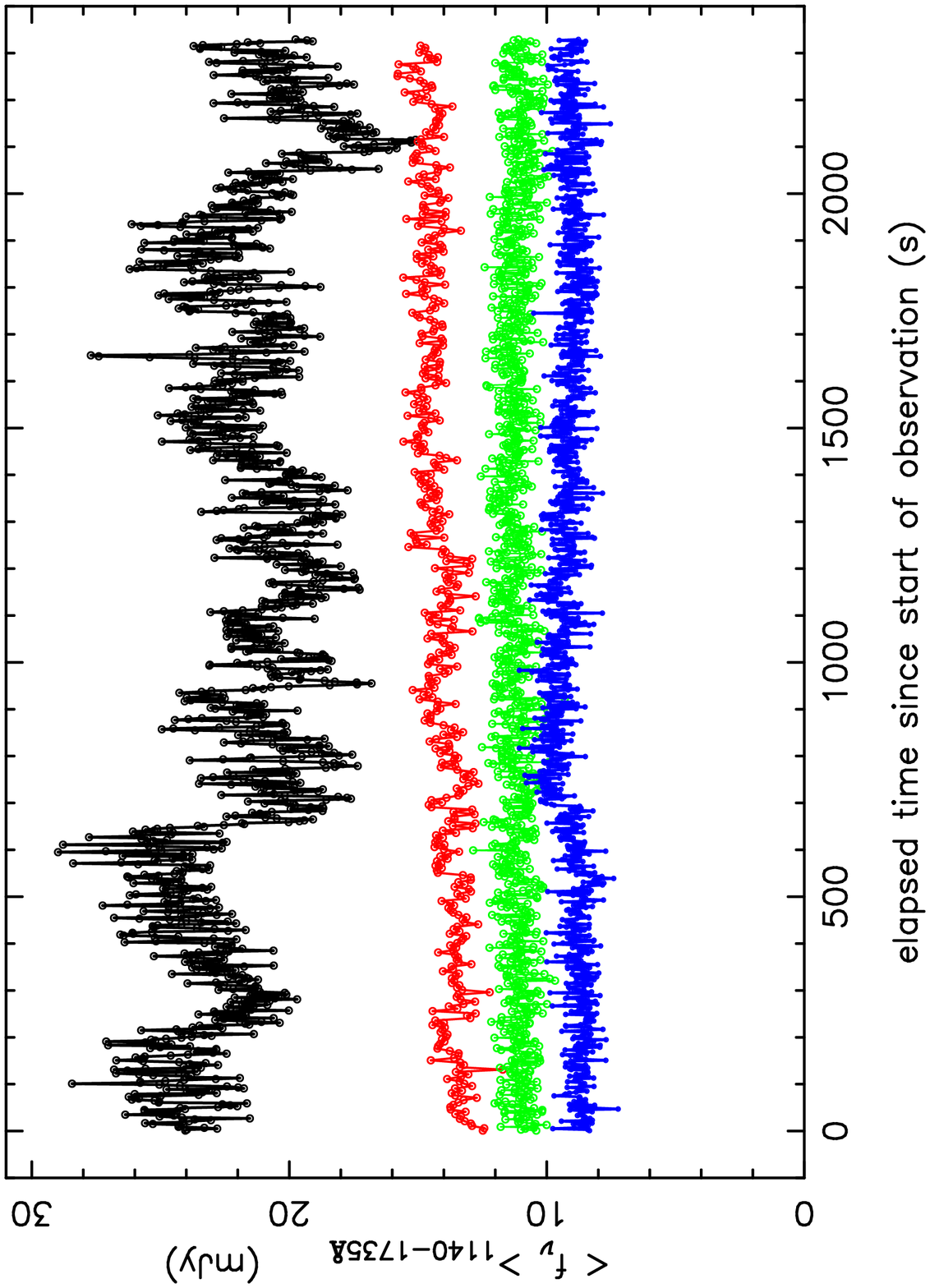}
\caption{ 
From top to bottom: Light curves of WZ~Sge obtained on UT~2001 Sep~11, 
Oct~10, Nov~10 and Dec~11 in the bandpass 1140--1735 \AA.
No offset in flux has been applied.
Notice the much larger amplitude flickering present on Sep 11.
\label{fig1}}
\end{figure}

\clearpage 
\begin{figure}
\figurenum{2}
\epsscale{0.8}
\plotone{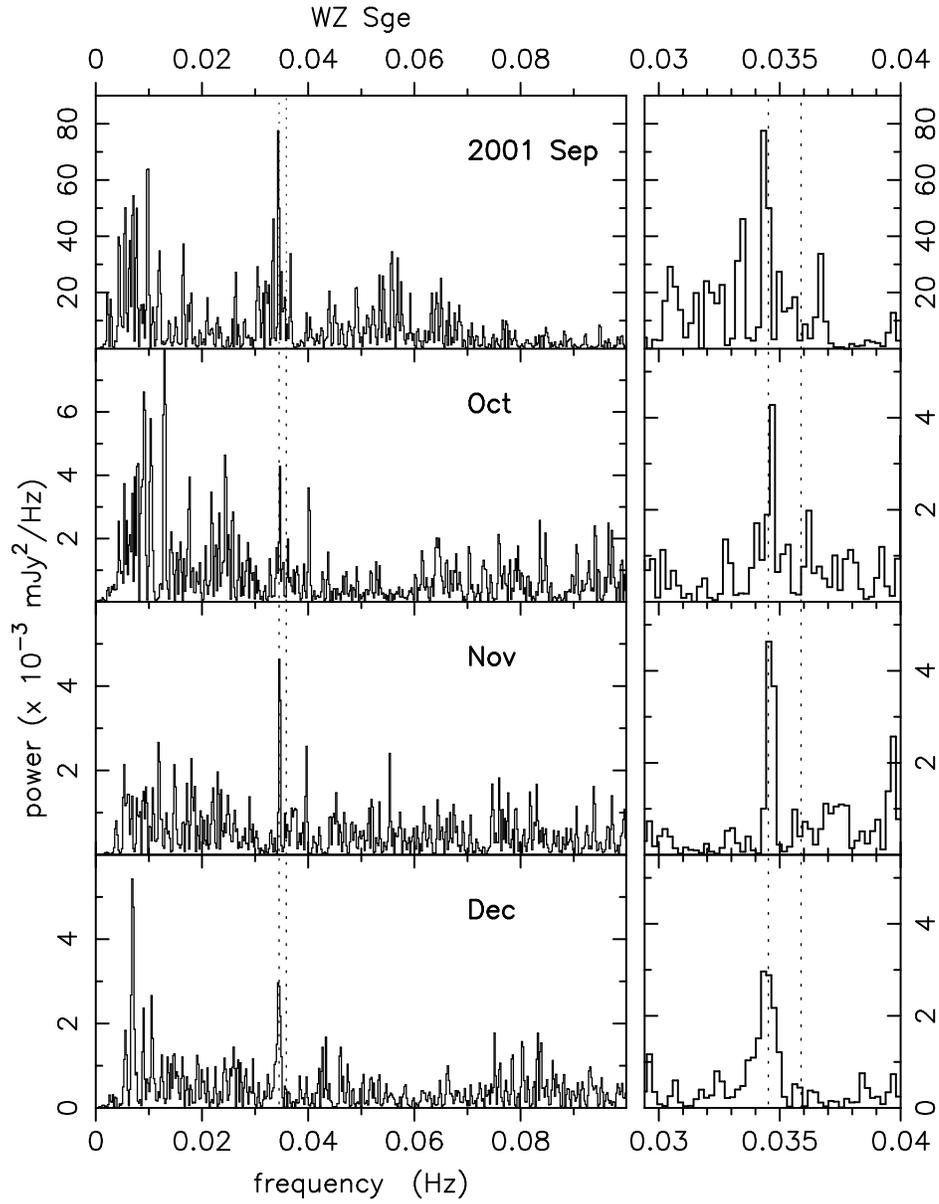}
\caption{ The low--frequency part of the power spectra of the four light
curves shown in Fig.~1. 
The dotted vertical lines mark the previously well--known 27.87~s and
28.96~s periods. The right--hand panels show a close--up  of the region
between 25 and 34~s. The P29 periodicity is clearly present at all epochs.
\label{fig2}}
\end{figure}

\clearpage 
\begin{figure}
\figurenum{3}
\epsscale{0.8}
\plotone{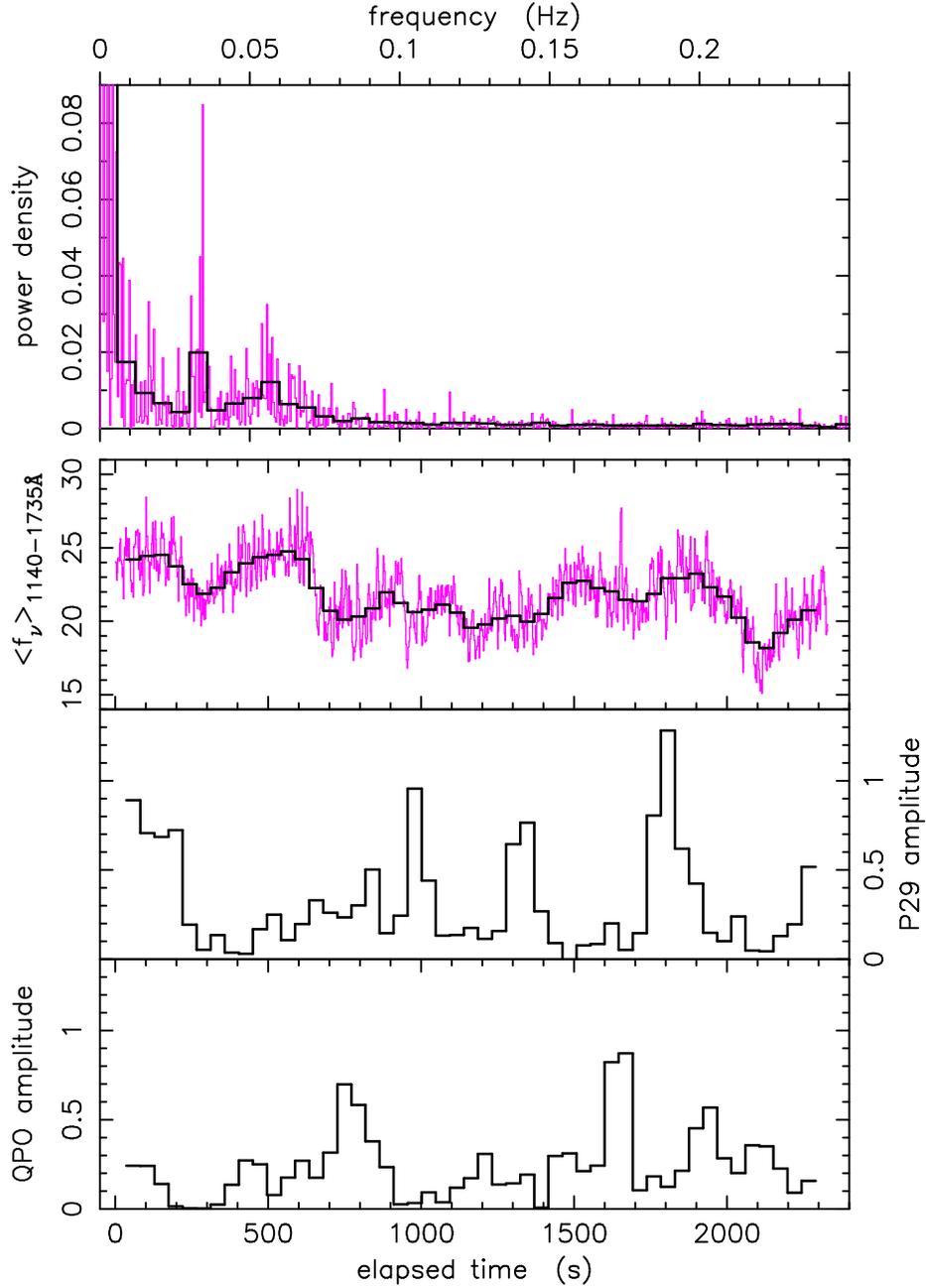}
\caption{ {\it Upper Panel:} The full power spectrum of the 2001 Sep light
curve after removing a linear trend (units are mJy$^{2}$/Hz).
Also shown is the same power spectrum after binning in frequency by 14:1.
The excess power due to the QPO near $\sim$ 18~s is readily visible
between 0.05 and 0.06 Hz. \ 
{\it Lower Panels, top to bottom:} The 2001 Sep mean light curve and
the light curve in 50 non--independent bins of duration 118~s. The two
lower panels show the highly variable amplitude of the P29 oscillation
and the QPO (units are in mJy). There is no obvious correlation between
the mean light curve and the periodicities, nor between the P29 and QPO
periodicities.
\label{fig3}}
\end{figure}

\clearpage 
\begin{deluxetable}{ccrcccc}
\tabletypesize{\scriptsize}
\tablecolumns{7}
\tablewidth{0pt}
\tablecaption{29~s Oscillations in WZ~Sge During Outburst \label{tbl-1}}
\tablehead{
\colhead{Date} & \colhead{Start Time} & \colhead{Mean UV Flux}  &
\colhead{Best--fit} & \colhead{Best--fit} & \colhead{Relative} &
\colhead{Unbiased 28.96 s}\\
\colhead{(yyyy-mm--dd)} & \colhead{(UT)}  & \colhead{(mJy)}  &
\colhead{Period (s)}    & \colhead{Amplitude (mJy)} &
\colhead{Amplitude (\%)} &
\colhead{Amplitude (mJy)}
}
\startdata
2001--09--11 & 03:03:28 & 21.8 $\pm$2.2 & 29.068 $\pm$0.067 & 0.416
$\pm$0.021 & 1.9 $\pm$0.2 & 0.339\\
2001--10--10 & 12:57:13 & 14.1 $\pm$0.6 & 28.844 $\pm$0.085 & 0.093
$\pm$0.018 & 0.7 $\pm$0.1 & 0.074\\
2001--11--10 & 11:45:32 & 11.2 $\pm$0.4 & 28.877 $\pm$0.053 & 0.102
$\pm$0.015 & 0.9 $\pm$0.1 & 0.093\\
2001--12--11 & 00:34:59 &  9.1 $\pm$0.5 & 29.052 $\pm$0.111 & 0.079
$\pm$0.014 & 0.9 $\pm$0.2 & 0.077\\

\tablecomments{Unbiased amplitudes are determined by holding the period
fixed at 28.96 s in the fit; uncertainties are the same as in the
least--squares best--fit sinusoid case. 
In all cases, the formal period resolution at 29 s is 0.36 s.
}
\enddata
\end{deluxetable}

\end{document}